\documentclass{article}

\usepackage{geometry}
\usepackage{float}
\usepackage{graphicx}

\usepackage{epstopdf}
\usepackage{subfigure}

\usepackage{natbib}
\bibpunct[, ]{(}{)}{;}{a}{}{,}

\usepackage{amsmath}
\usepackage{amsfonts}
\usepackage{amsthm}

\theoremstyle{plain}

\theoremstyle{definition}

\theoremstyle{remark}

\def\mgm3{$\mu g\,/\,m^3$}

\begin{document}


\title{Debiasing physico-chemical models in air quality monitoring by combining different pollutant concentration measures}


\author{Benjamin Auder\textsuperscript{a}, Camille Coron,\textsuperscript{b}
Jean-Michel Poggi\textsuperscript{a,c}, Emma Thulliez\textsuperscript{d}}

\date{}

\maketitle
\noindent%
\textsuperscript{a} LMO, University Paris-Saclay, bat. 307, Rue Michel Magat,
91400 Orsay, France; \\
\textsuperscript{b} MIA-Paris-Saclay, INRAE,  AgroParisTech, Palaiseau,
France;\\
\textsuperscript{c} IUT Paris, dept SD, University Paris Cité, Paris, France;\\
\textsuperscript{d} INSA Rouen Normandie, Normandie Univ, LMI UR 3226, F-76000 Rouen, France

\abstract{
Air quality monitoring requires to produce accurate estimation of nitrogen dioxide or fine particulate matter concentration maps, at different moments. A typical strategy is to combine different types of data. On the one hand, concentration maps produced by deterministic physicochemical models at urban scale, and on the other hand, concentration measures made at different points, different moments, and by different devices. These measures are provided first by a small number of reference stations, which give reliable measurements of the concentration, and second by a larger number of micro-sensors, which give biased and noisier measurements. 

The proposed approach consists in modeling the bias of the physicochemical model and estimating the parameters of this bias using all the available concentration measures. Our model relies on a partition of the geographical space of interest into different zones within which the bias is assumed to be modeled by a single affine transformation of the actual concentration. Our approach allows to improve the concentration maps provided by the deterministic models but also to understand the behavior of micro-sensors and their contribution in improving air quality monitoring. 

We introduce the model, detail its implementation and experiment it through numerical results using datasets collected in Grenoble (France).}


\vspace{1ex}




\section{Introduction}

In air quality monitoring, providing high-quality estimation of the concentration of some pollutant, nitrogen dioxide (NO$_2$) or particulate matter ($PM$), for example, is mandatory. To this aim, organizations can use measurements of concentration. We focus here on a situation where it is assumed that two types of spatio-temporal concentration measurement data coexist: network 1 provides high quality measurements, taken at a regular frequency and at fixed points; network 2 provides lower quality measurements, which suffer from biases that may change over time. This situation has become quite common, as those kind of network has been studied at country \cite{huang2018real} or city scale \cite{Gressentetal2020}. We focus here on the urban scale.

Organizations can also make use of physicochemical models. Their outputs classically provide an initial map which is then corrected using measurements by assimilation/geostatistical techniques. A way to improve the quality of reconstructed maps is to increase the density of sensors. Availability of micro-sensors in addition to reference stations measurements opens a possibility for this improvement, without a prohibitive cost. This has opened a collection of papers dealing with various aspects illustrating the interest of such new sensors and providing technical hints to properly incorporate them in the data collection and analysis process.

Conventional approaches to improving pollution maps involve the use of geostatistical methods to correct numerical model outputs with measurements from reference stations. These approaches use data assimilation methods (see, for example \cite{de2011geostatistical}), and the added value of this correction of model outputs with observations is well established. In recent years, the availability of low-cost sensors to measure pollutant concentrations in addition to measurements from reference stations means that these concentration maps can be further improved without prohibitive cost. A very interesting study of such an approach \cite{schneider2017mapping} looks at nitrogen dioxide concentration maps in Oslo (Norway). A similar study \cite{Gressentetal2020} has been carried out for Nantes (France), including data from mobile micro-sensors. More recently, the design of hierarchical networks, low-cost sensor verification and correction by data fusion in these networks have also been presented in \cite{miskell2018solution} or in \cite{weissert2020hierarchical}. Let us mention also an iterative correction method proposed recently 
\cite{bobbia2022spatial}. All these approaches are interesting, but they are based on raw or corrected data that are merged with maps, and the whole is not integrated (the same corrected data would be used regardless of the model used to generate the map). 

More generally, an important literature on spatiotemporal statistical models for calibration \cite{huang2018real} and data fusion \cite{gilani2019nonstationary} \cite{ma2020hierarchical}) is available and related methods are of interest.

An alternative approach \cite{giraud2016capitalizing,coron2018bayesian} to merging data from different protocols has recently been explored in the context of biodiversity monitoring. In this question, the authors have two datasets of bird observations made by ornithologists: one is the result of observations made at fixed locations and for an imposed duration; the other is the result of observations made at non-prescribed locations and for an unknown duration. The first set of data provides important information and makes it possible to produce maps of species abundance. The second dataset cannot be used on its own due to the lack of an observation protocol, but contains a very large number of observations. The idea of the approach proposed by the authors is to calibrate the second dataset using the first, in order to benefit from both the information in the first dataset and the abundance in the second. Thanks to this combination of datasets, the authors are able to provide more accurate abundance maps for the different species under consideration. This seminal idea was one of the starting points of this work.

The proposed approach inverts the traditional perspective in statistics by starting from the physicochemical model and modeling the bias to finally correct the output map. This approach makes it possible, on the one hand, to lean on the physicochemical model definition to model its potential errors that can be specific to the area of interest, and on the other hand, to get an interpretation of the distances between model outputs and measurements, with the potential aim of improving not only concentration maps but also physicochemical models on the considered geographical area.

For sure, spatial correlation and temporal correlation are key issues in the traditional approach to model the dynamics of the pollutant. However, our approach in this work consists in assuming that the main spatio-temporal dynamics of the pollutant (notably due to pollutant emissions and urban geography) are already taken into account by the deterministic model and that they are not present in its bias. Then, we suppose that a considerably simpler model can be useful, at least as a starting point. Finally, if some spatio-temporal patterns remain, the addition in the future of other covariates to explain the bias could be considered.

Our work deals with air quality monitoring, by combining different types of data. More precisely, our aim is to produce (typically at the scale of a large given city) accurate estimation of nitrogen dioxide or fine particulate matter concentration maps, at different moments. Therefore, we have at our disposal, on the one hand, concentration maps produced by deterministic physicochemical models (such as SIRANE \cite{SIRANE-1,SIRANE-2} or ADMS-urban \cite{mchugh1997adms,carruthers1998development}), and on the other hand, concentration measures made at different points, different moments, and by different devices. These measures are provided first by a small number of reference stations, which give reliable measurements of the concentration, and second by a larger number of micro-sensors, which give biased and noisier measurements.

The proposed approach consists in modeling the bias of the physicochemical model (e.g. due to model assumptions that are not satisfied in practice, such as constant altitude) and estimating the parameters of this bias using all the available concentration measures. Our model relies on a partition of the geographical space of interest into different zones within which the bias is assumed to be modeled by a single affine transformation of the actual concentration. Our approach allows to improve the concentration maps provided by the deterministic models but also to understand the behavior of micro-sensors and their contribution in improving air quality monitoring. We will introduce the model, detail its implementation and experiment it through numerical results using datasets collected on the Grenoble area (France).

The paper is organized as follows. Section 2 provides our model and approach. Section 3 illustrates some aspects of the method, including numerical results, by applying it to a real dataset coming from the Grenoble area, in France. Finally, Section 4 opens some perspectives.


\section{The model}

We propose a simple toy model, which in particular does not include meteorological data. This model is inspired by papers \cite{soulhac2017model, beauchamp2018additive}, in which the authors analyse the sources of error in SIRANE type maps. We initially identified three sources of error: Firstly, the physical models used to produce the SIRANE maps assume a constant altitude, which is a major source of error, particularly for areas surrounded by hills (such as those found in Grenoble); secondly, these models seem to perform differently depending on the type of area (urban, suburban, rural); finally, in \cite{soulhac2017model}, the authors mention that the bias of the SIRANE maps seems to increase with concentration (which may be partly related to the previous point).

Our model, described in the following subsections, then aims at modeling the behaviour of  micro-sensors, reference stations and physico-chemical models. It should be noted that for the moment we are not yet taking into account seasonality, which is an important element in the structuring of the bias present in the maps derived from physico-chemical models. However we will consider that the behavior of micro-sensors and physico-chemical models depend on the hour of the day. Incorporating more temporal aspects into this bias modeling will be an objective in the future, for example by taking inspiration from \cite{fasso2007air}. Several avenues for improvement requiring a more sophisticated model are mentioned in the last section.

\subsection{Data modeling}

We denote by $P_s(t)$ the concentration of the pollutant that we are trying to estimate at a point $s\in\mathbb{R}^2$ and at time $t$. We assume that we have access to measures of the concentration of this pollutant realized by $I$ reference stations. We assume that for each $i$ in $\{1,2,...,I\}$ the reference station $i$ is located at $s_{i,1}\in\mathbb{R}^2$ (assumed to be fixed in this work) and we denote by $X^{i,1}(t)$ the measurement provided by this reference station at time $t$. The measurements $X^{i,1}(t)$ are then assumed to follow the model:
\begin{equation}
X^{i,1}(t)=P_{s_{i,1}}(t)+\epsilon^{(1)}_{s_{i,1}}(t),
\label{sensors}
\end{equation}
where $\epsilon^{(1)}_{s_{i,1}}(t)$ are independent random variables, centered and normally distributed with unknown variance $\sigma_{i,1}^2$. In other words, the true pollution is observed corrupted by an additive noise, assumed to be centered. Note that it is not necessary for the noise model to be a Gaussian distribution; this assumption will only play a role when inferring the parameters and a generalization can be proposed.

Now we assume that we have access to measures of the concentration of the same pollutant realized by $J$ micro-sensors. We assume that for each $j$ in $\{1,2,...,J\}$ the micro-sensor $j$ is located at $s_{j,2}\in\mathbb{R}^2$ (also assumed to be fixed in this work) and we denote by $X^{j,2}(t)$ the measurement provided by this reference station at time $t$. The measurements $X^{j,2}(t)$ are assumed to be modeled by
\begin{equation}
X^{j,2}(t)=\alpha_j P_{s_{j,2}}(t) + \beta_j + \epsilon^{(2)}_{s_{j,2}}(t),
\label{micro_sensors}
\end{equation}
where $\epsilon^{(2)}_{s_{j,2}}(t)$ are independent centered Gaussian random variables with unknown variance $\sigma_{j,2}^2$.
The locations of the micro-sensors are imposed and generally do not coincide with those of the reference stations. The parameters $\alpha_j$ and $\beta_j$ in Equation \eqref{micro_sensors} can either be assumed to be constant or to depend on the hour of the day, on the season, or on the month of the year. In particular, the last assumption is a direct way of checking the evolution of the micro-sensors' behaviour over time.

\subsection{Physico-chemical model and bias modeling}

We also assume that we have a map derived from physico-chemical models of the SIRANE type, which gives us an initial prediction $\widetilde{P}$ of the concentration $P$ at any point $s$ (in particular at the micro-sensor and reference stations locations) and at any time $t$. The link between $\widetilde{P}$ and $P$ is modeled by

\begin{equation}
\widetilde{P_s}(t)=  P_{s}(t)+ B_s(t),
\label{prediction}
\end{equation} 
where the term $B_s(t)$ is therefore the bias of the $\widetilde{P}$ map that we are trying to understand and estimate. Our approach consists in modeling this bias and estimating the parameters of this model from the sets of data $X^{i,1}(t)$ and $X^{j,2}(t)$.

In this work we will focus on a situation in which the considered space can be divided in different zones, on which the bias is an affine function of the true concentration. This affine dependence is inspired from observations made in \cite{SIRANE-1}. More precisely we assume that the bias satisfies
\begin{equation}
\label{bias}
B_{s}(t)=C(s)+\rho(s) P_{s}(t)
\end{equation}

where $C$ and $\rho$ are two piecewise constant functions whose zones of constancy are the same and will have to be determined in discussion with the data suppliers or according to a definition that can be implemented automatically (for example type of zone, rural, urban or peri-urban). 
We will then denote by $A_1,...,A_K$ these different zones, and set $C(s)=\sum_{k=1}^KC_k\mathbf{1}_{A_k}(s)$ and $\rho(s)=\sum_{k=1}^K\rho_k\mathbf{1}_{A_k}(s)$. The number $K$ is then the number of zones on which these functions are constant, and $C_k$ (resp. $\rho_k$) is the value of the function $C$ (resp. $\rho$) on the zone $A_k$. 

The function $C$ models the bias of the map in the absence of concentration at this point, while the function $\rho$ models the dependence of the error of the initial map on the particle concentration. Finally, for identifiability reasons that will be explained in the next section, the zones on which $C$ and $\rho$ are constant are defined so that each block contains at least one reference station.

\subsection{Parameter estimation} 
\subsubsection{Formulation of the estimation problem}
Combining the two last equations, we have $$B_{s}(t)=C(s)+\rho(s) (\widetilde{P_{s}}(t)-B_{s}(t))$$
and then
$$B_s(t)=\frac{C(s)+\rho(s) (\widetilde{P_s}(t))}{1+\rho(s)}.$$

Considering first the reference stations measures gives

\begin{align*}X^{i,1}(t)&=P_{s_{i,1}}(t)+\epsilon^{(1)}_{s_{i,1}}(t)\\&
=\widetilde{P}_{s_{i,1}}(t)-B_{s_{i,1}}(t)+\epsilon^{(1)}_{s_{i,1}}(t)\\&
=\widetilde{P}_{s_{i,1}}(t)-\frac{C(s_{i,1})+\rho(s_{i,1}) \widetilde{P}_{s_{i,1}}(t)}{1+\rho(s_{i,1})}+\epsilon^{(1)}_{s_{i,1}}(t),
\end{align*} giving

\begin{align}\label{eq:stations}X^{i,1}(t)=\frac{1}{1+\rho(s_{i,1})}\widetilde{P}_{s_{i,1}}(t)-\frac{C(s_{i,1})}{1+\rho(s_{i,1})}+\epsilon^{(1)}_{s_{i,1}}(t).
\end{align}

It turns out that the observed concentrations are modeled as a function of the output from the physical models. It could be reasonable to assume that these models also somehow take into account actual measurements from stations, but this is not the case for SIRANE. In addition, the dependence is not explicit at all and then it would be reasonable to consider the equations conditionally to the model outputs.

Similarly, by considering the micro-sensors measures

 \begin{align*}X^{j,2}(t)&=\alpha_j P_{s_{j,2}}(t)+\beta_j+\epsilon^{(2)}_{s_{j,2}}(t)\\&
=\alpha_j(\widetilde{P}_{s_{j,2}}(t)-B_{s_{j,2}}(t))+\beta_j+\epsilon^{(2)}_{s_{j,2}}(t)\\&
=\alpha_j\left(\widetilde{P}_{s_{j,2}}(t)-\frac{C(s_{j,2})+\rho(s_{j,2}) \widetilde{P}_{s_{j,2}}(t)}{1+\rho(s_{j,2})}\right)+\beta_j+\epsilon^{(2)}_{s_{j,2}}(t)
\end{align*} 

then 

\begin{equation}\label{eq:MC}X^{j,2}(t)=\frac{\alpha_j}{1+\rho(s_{j,2})}\widetilde{P}_{s_{j,2}}(t)-\frac{\alpha_j}{1+\rho(s_{j,2})}C(s_{j,2})+\beta_j+\epsilon^{(2)}_{s_{j,2}}(t).\end{equation}

Recall that the functions $C$ and $\rho$ are piecewise constant functions, and the zones on which they are constant are assumed to be known, and the same for the two functions. The parameters that we wish to estimate from Equations \eqref{eq:stations} and \eqref{eq:MC} are first the values $(C_k)_{1\leq k\leq K}$ and $(\rho_k)_{1\leq k\leq K}$ of the piecewise constant functions $C$ and $\rho$ on each zone $(A_k)_{1\leq k\leq K}$, second, the transformation parameters $(\alpha_j,\beta_j)$ of each micro-sensor, and third the variances $\sigma_{i,1}^2$ and $\sigma_{j,2}^2$ for all $i\in\{1,...,I\}$ and $j\in\{1,...,J\}$ . If we assume that the variances only depend on the zone, i.e. if $\sigma_{i,1}^2=\sum_{k=1}^K\sigma(k)^2\mathbf{1}_{A_k}(s_{i,1})$ and $\sigma_{j,2}^2=\sum_{k=1}^K\sigma(k)^2\mathbf{1}_{A_k}(s_{j,2})$  where $\sigma(k)^2$ models the variance associated to the zone $A_k$, then the estimation of the $2K+2J$ parameters $((C_k)_{1\leq k\leq K}, (\rho_k)_{1\leq k\leq K}, (\alpha_j,\beta_j)_{1\leq j\leq J})$ as well as the variance parameters $(\sigma(k)^2)_{1\leq k\leq K}$ can be done using a linear modeling that we detail now.

Using data from the reference stations and from the micro-sensors, these equations can be used to estimate the parameters $\rho$ and $C$ and the parameters $\alpha_j$ and $\beta_j$ for each micro-sensor.

\subsubsection{Estimation procedure}

The combination of Equations \eqref{eq:stations} and \eqref{eq:MC} can be written as a linear model, that can be implemented to provide estimation of all parameters, separately for each zone. Indeed, let us denote by $J(k)$ the number of micro-sensors in zone $k$ and $\mathbf{X}=(X_1,...X_\mathcal{O})$ the vector of all measures collected by both reference stations and micro-sensors in zone $k$, $(\tilde{P}_1,...\tilde{P}_\mathcal{O})$ the vector of corresponding model outputs at the same time and position, and let us denote by $$\Theta=\left(\frac{1}{1+\rho_k},\frac{C_k}{1+\rho_k},\frac{\alpha^{(k)}_1}{1+\rho_k},\frac{\alpha^{(k)}_2}{1+\rho_k},...,\frac{\alpha^{(j)}_{J(k)}}{1+\rho_k},\beta^{(k)}_1-\frac{\alpha^{(k)}_1C_k}{1+\rho_k},...,\beta^{(k)}_{J(k)}-\frac{\alpha^{(k)}_{J(k)}C_k}{1+\rho_k}\right)$$ the vector of regression parameters that we wish to estimate, where $(\alpha_i^{(k)},\beta_i^{(k)})$ are the transformation parameters of the $i$-th micro-sensor in zone $k$, introduced in Equation \eqref{micro_sensors}. Then from Equations \eqref{eq:stations} and \eqref{eq:MC}, \begin{equation}\label{eq:LinearModel}\mathbf{X}=A\Theta+\mathbf{\varepsilon},\end{equation} where $\mathbf{\varepsilon}$ is a random centered vector whose entries are independent, with variance $\sigma(k)^2$, and the matrix $A$ is defined as follows. If the $l$-th observation is made by a reference station, then $$A_{l\cdot}=(\tilde{P}_{l},-1,(0,...,0),(0,...,0))$$ where the last two vectors are null vectors with dimension $J(k)$. If the $l$-th observation is made by the micro-sensor numbered by $j$, then
$$A_{l\cdot}=(0,0,(0,...,\tilde{P}_{l},0,...0),(0,...,1,0,...,0)),$$ where the last two vectors have all null quantities except in position $j$. 
Since there is at least one reference station in each zone, the linear modeling proposed in Equation \eqref{eq:LinearModel} allows to estimate $\frac{1}{1+\rho_k}$ and $\frac{C_k}{1+\rho_k}$ for all $k$ in $1,...,K$, and $\beta_j$ and $\frac{\alpha_j}{1+\rho(s_{j,2})}$ for all $j$ in $1,...,J$, by computing the least-square estimator:
$$\hat{\Theta}=\arg\min_{\theta} \|\mathbf{X}-A\theta\|^2.$$
This estimation procedure also gives an estimation for the variance parameter $\sigma(k)^2$ in each zone $k$.

It should be noted that micro-sensors whose position does not lie in the space of definition of the $C$ and $\rho$ functions (constrained by the necessary presence of at least one reference station in each zone of constancy) cannot be used to infer the model parameters (at least as long as their $\alpha_i$ and $\beta_i$ parameters are not assumed to be known). If necessary, they are omitted.

\subsection{Alternative concentration maps}

A second natural approach consists in estimating first the parameters $(C_k)_{1\leq k\leq K}$, $(\rho_k)_{1\leq k\leq K}$, and $\sigma(k)^2$ using reference stations only. Indeed Equation \eqref{eq:stations} gives a linear relationship between measures and model outputs, which provides estimates of all these parameters, since it gives estimates for $\frac{1}{1+\rho_k}$, for $\frac{C_k}{1+\rho_k}$, and for $\frac{\sigma(k)^2}{1+\rho_k}$ for all $k$ in $1,..,K$. This will give a new corrected map, that is obtained using only reference stations. Comparing this corrected map to the one obtained previously using both reference stations and micro-sensors can then help understanding and measuring the benefit of combining micro-sensors and reference stations measurements, to improve concentration maps.

Next, the parameters $(\alpha_j,\beta_j)_{1\leq j\leq J}$ can be estimated using only micro-sensors measures, using the linear relationship between micro-sensors measures and model outputs given by Equation \eqref{eq:MC}, and setting all the previously estimated parameters to their respective estimated values. This approach will however result in a separate estimation of the variance of micro-sensors measure, i.e. $\sigma_{i,1}^2=\sigma_1(k)^2$ while $\sigma_{j,2}^2=\sigma_2(k)^2$. In that case note that micro-sensors are not used to reconstruct the corrected map, but on the contrary, in some sense, the corrected map helps understanding the behavior of micro-sensors. 

Finally, the estimation of the parameters $(\alpha_j,\beta_j)_{1\leq j\leq J}$ provides corrected micro-sensors measures that can now considered as reference stations. This allows to consider now that space is divided into a larger number of zones (as many zones as devices) on which the functions $C$ and $\rho$ are constant, and to provide a new concentration maps in which the estimated bias has a finer resolution.

To sum up, our procedure produces then three concentration maps: the first one is obtained by estimating parameters of the model by pooling all measures made by both reference stations and micro-sensors (sometimes referred to as "pool" in the sequel); the second one is obtained by using only reference stations measures (referred to as "no\_ms"); the last one is obtained by correcting micro-sensors using fixed stations, and considering them as reference stations in a second time (referred to as "ms\_as\_sta").

\section{A real data example}

\subsection{Measurements data}

ATMO Normandie, the organization in charge of air quality monitoring in Normandie including the city of Rouen, uses SIRANE to evaluate air pollution at urban scale. It collects measurements from a few monitoring stations and numerous micro-sensors. At this stage, the data required for this study is not fully available. Consequently, we will illustrate the method using data from another organization responsible of air quality monitoring in France: ATMO Auvergne-Rhône-Alpes, whose area covers the city of Grenoble.

The data used in this study is part of the Mobicit'air project of Atmo Auvergne-Rhône-Alpes, the organization supervising air quality in Grenoble, in partnership with Grenoble-Alpes-Métropole mainly concentrated in the city.
 
\begin{figure}[H]
\centering
\includegraphics[scale=0.35]{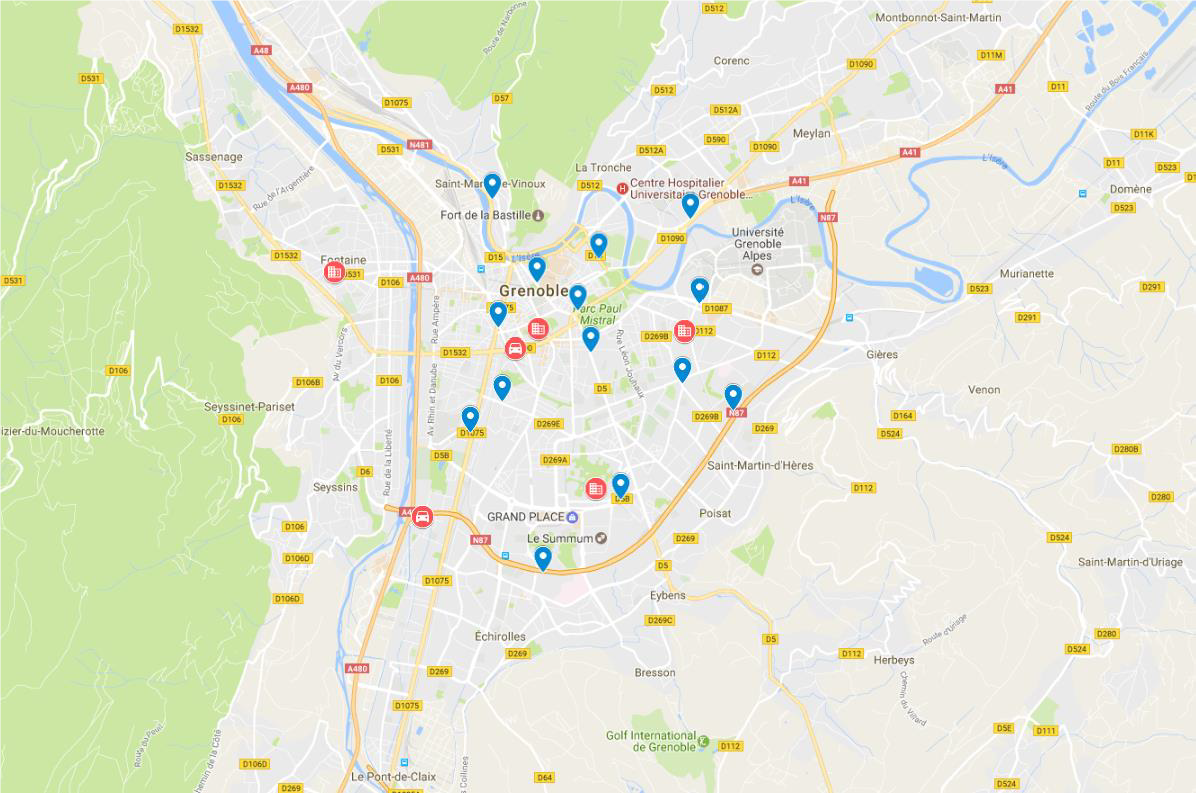}
\caption{Location of the sites in the Grenoble agglomeration, the sites in blue correspond to the micro-sensors, the sites in red correspond to the reference sites. This figure is extracted from \circledR Atmo Auvergne-Rhône-Alpes (2017) \textit{ Assimilation de données de micro-capteurs dans les cartographies fines échelles}.}
\label{fig:CarteGrenoble_origine} 
\end{figure}

It consists of hourly average concentrations (in \mgm3) of NO$_2$ pollutant
during 147 days, from January 5, 2017 to May 31, 2017. We focus on the data
corresponding to the months of January and February 2017, a period of the year during which NO$_2$ concentrations are the highest. The device network is composed of 9 reference stations located in the Grenoble area, close to traffic, in urban or suburban places and of 12 micro-sensors. Devices sites from both networks are presented in Figure \ref{fig:CarteGrenoble_origine}. One can notice that three monitoring stations are far from Grenoble downtown, as shown in Figure \ref{fig:C_deux_idees}, and therefore are not present on the close-up.

For a more detailed presentation of the data, we refer to the unpublished report publicly available online \cite{AtmoGA1}. Besides, note that this dataset was also studied in \cite{bobbia2022spatial} to propose a spatial correction method of sensor measurements, which is a related but different problem.

\subsection{Deterministic model outputs}

In addition, we have two maps that can be used as deterministic maps: a SIRANE map averaged over the year 2017 with a resolution of 10 meters (we call it $(Sirane(s))$, and a CHIMERE map averaged over every hour between January and February 2017 with a resolution of 3 km, called $(Chimere(s,t))$. It should be noted that this CHIMERE map is taken to allow quick computations and reduce the cost of storing SIRANE maps at every hour. With these two maps, we have so far considered different options for the initial deterministic map aimed to be corrected by measures: either one of the two maps, or a combination of the two maps. The best option that we found, that will be considered from now on, is to take the average of the two available maps as initial deterministic map:
\begin{equation}\label{eq:cartemoyenne}\widetilde{P_s}(t)=1/2\times (Sirane(s)+Chimere(s,t)).\end{equation}

The main interest of considering the mean of the two maps is to benefit both from the temporal variability brought by the CHIMERE map, and from the fine spatial resolution of the SIRANE map.

\subsection{Adapting the model}

\subsubsection{Zone definition}

\begin{figure}[H]
  \begin{minipage}[b]{0.45\linewidth}
   \centering
   \includegraphics[width=6cm,height=7cm]{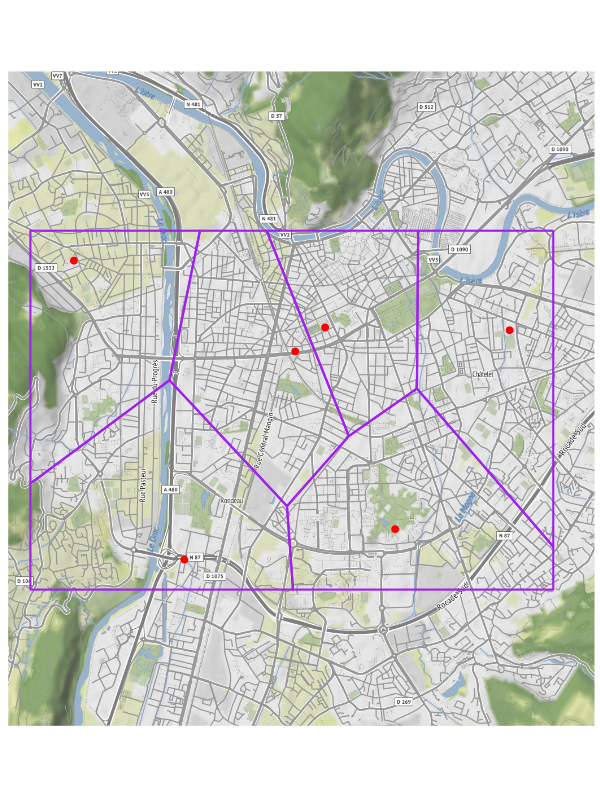}     
  \end{minipage}
\hfill
  \begin{minipage}[b]{0.45\linewidth}
   \centering
   \includegraphics[width=6cm,height=7cm]{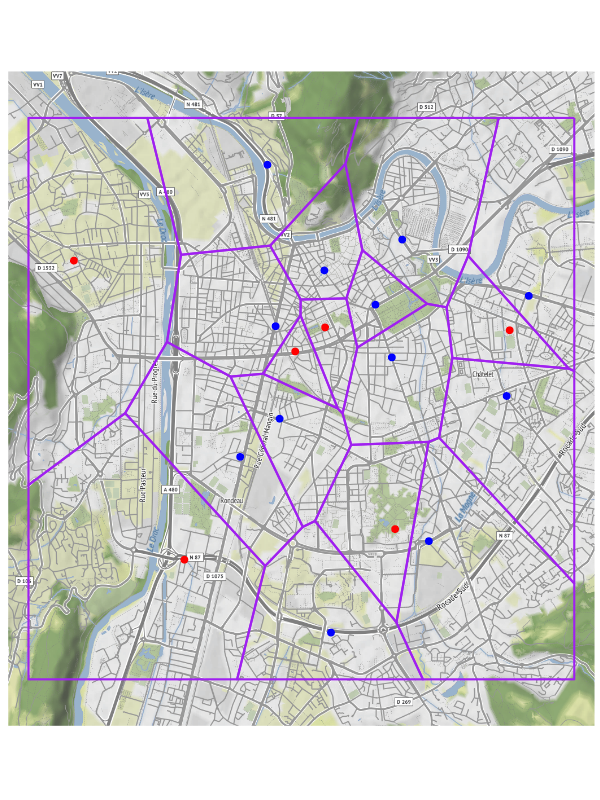}     
  \end{minipage}

  \caption{Two illustrations of the division into different zones defining the $C$ and $\rho$ functions. On the left: considering only the reference stations; on the right: considering the whole network. Each block contains exactly one device.}
  \label{fig:C_deux_idees}

\end{figure}

We now define the zones of constancy for the functions $C$ and $ \rho$, bearing in mind that each zone must include at least one reference station. Figure \ref{fig:C_deux_idees} shows how these zones are defined, at two different scales: on the left only reference stations are considered, while on the right microsensors are used too (blue dots). In both cases zones corresponds to Voronoi cells containing exactly one device each. The results presented in the following sections will consider and compare both approaches.

This partition is based on a theoretical definition and has two  disadvantages: firstly, it links micro-sensors and reference stations that are sometimes very far apart; secondly, the pollution within each zone thus formed is very heterogeneous. Subsequently, we plan in the future to propose an alternative division that could be defined using, for example, the average pollution level given by the $(P_s)_s$ map, or the distance to the main roads.

\subsubsection{Bias definition} 
We then define the functions $C$ and $\rho$ so that they are constant over each of the zones thus defined: $$C(s)=\sum_{i=1}^KC_i\mathbf{1}_{Z_i}(s),$$ and 

$$\rho(s)=\sum_{i=1}^Kr_i\mathbf{1}_{Z_i}(s),$$ where the parameters $C_i$ and $r_i$ will be estimated using the data and $K$ is the number of zones. 

\subsubsection{Modeling the data and estimating the parameters} 

We implement the linear model given by Equations \eqref{eq:stations} and \eqref{eq:MC}. This model allows us to estimate $\rho$ and $C$ and the $\alpha_j$ and $\beta_j$ from reference stations data and from micro-sensors data. It should be noted that these parameters can also depend on the hour of day without causing problems of identifiability. We have considered both the case where the parameters $\rho$, $C$, $\alpha_j$ and $\beta_j$ do not depend on time, and where these parameters do depend on time (in which case the inference is then simply made independently for each considered time slot). We only show the estimation results in the case where these parameters depend on the hour of the day, as they provide more relevant concentration maps.

As we focus on a short period of time ($2$ months in total), we neglect seasonality effets and decompose the considered period as follows: the learning period is in January and the testing period is in February. All estimations are made using measurements made during the learning period, and their performance are assessed by comparing the generated prediction to the real measurements made by reference stations during the testing period.

\section{Numerical results}

We describe and comment on a selection of the results we have obtained, although this work is still in progress, focusing on various aspects: spatial and temporal aspects of the corrected maps, the correction parameters and a first cross-validation error. 

\subsection{Correcting deterministic maps}

\begin{figure}[H]
  \centering
  \includegraphics[width=0.6\textwidth]{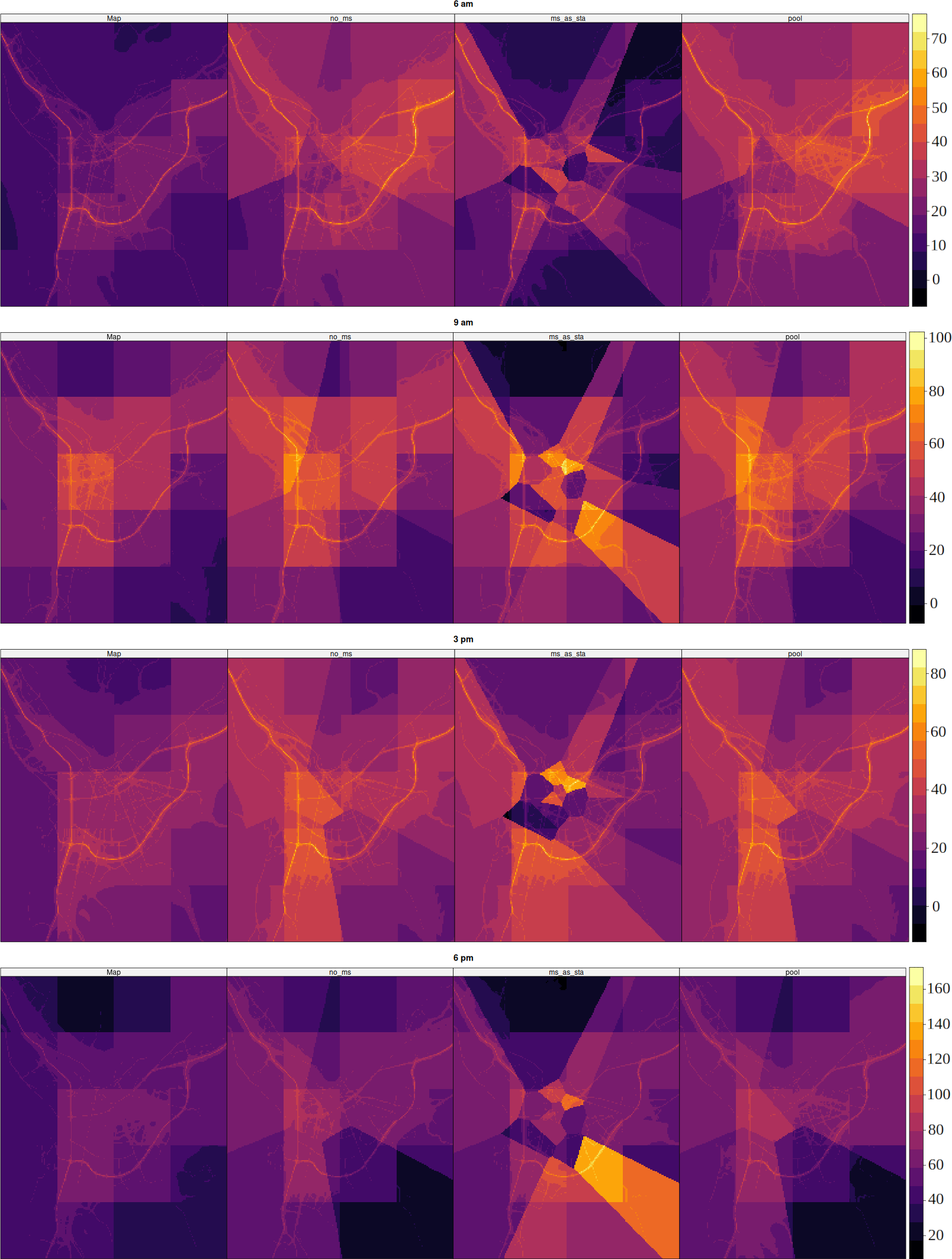}
  \caption{Concentration maps for Thursday 9 February 2017, at different times of day: on the left the initial map, and on the right the corrected maps (from left to right: no\_ms, ms\_as\_sta, pool). From top to bottom: 6 am, 9 am, 3 pm, 6 pm.\label{fig:moycarte_jour}}
\end{figure}

The first aim of our approach is to provide a corrected version of the initial maps. This correction is illustrated in Figure \ref{fig:moycarte_jour}: on the left the initial map and on the right the corrected maps using measurements and integrating our modeling approach on a given day: Thursday 9 February 2017. Here we consider four different hours: from the top to the bottom 6 am, 9 am, 3 pm and 6 pm. 

It should be noted that the corrected maps present higher average concentrations and enlighten more clearly the roads of the north-west and east of the town. The differences between the hours are consistent with what is expected. Indeed, we observe low signal during the night (6 am, 9 pm) and two peak hours with potentially high signal at 9 am and 6 pm. Moreover, a comparison between the corrected maps shows that the model using only reference stations (no\_ms) and the model using both devices type simultaneously (pool) provide similar bias estimations. That is mostly due to the fact that the zones are the same for both methods. However, the use of the micro-sensors in the pool model raises some difference, as it can be seen at 6 am, when concentration are higher near road with sensors than without. Finally, those two correction methods show great difference with the one using the corrected sensors as if they were reference stations (ms\_as\_sta). In that case, the zoning pattern appears more clearly, leading to possible over-estimations, as at 9 pm in the south-east of the city. Due to the lack of smoothness and the possible residual bias of the sensors, the zones are very visible on the corrected maps even though they have no obvious meaning in terms of pollution.

\begin{figure}[H]
\centering
   \includegraphics[width=0.8\textwidth]{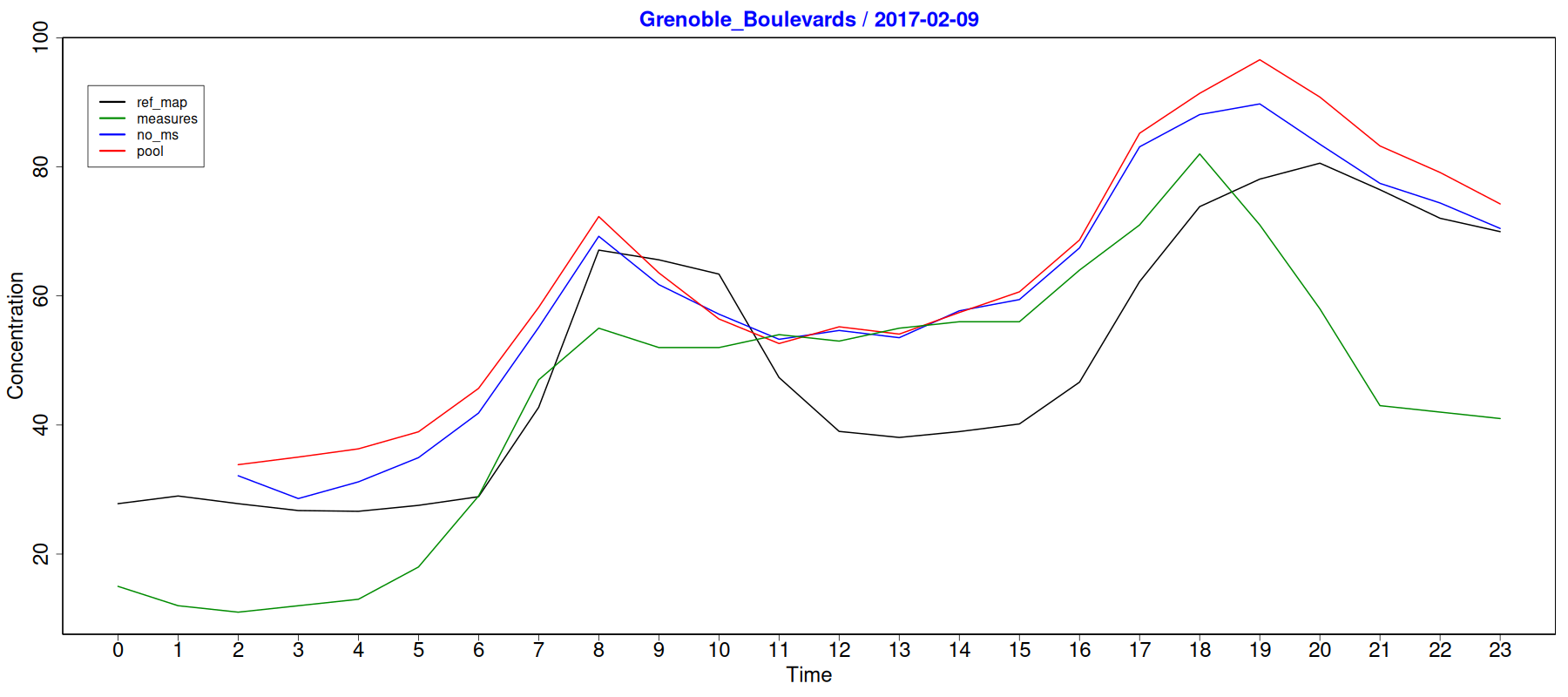}     
  \caption{Concentrations for the area corresponding to the Grenoble Boulevards reference station, for 9 February 2017, as a function of time. In black: the initial map; in green: measurements from the Grenoble Boulevards reference station;
  in red: estimated concentration by pooling all data; in blue: estimated concentration using only reference stations.}
  \label{fig:gre_bd_jour}
\end{figure}

\begin{figure}[H]
\centering
   \includegraphics[width=\textwidth]{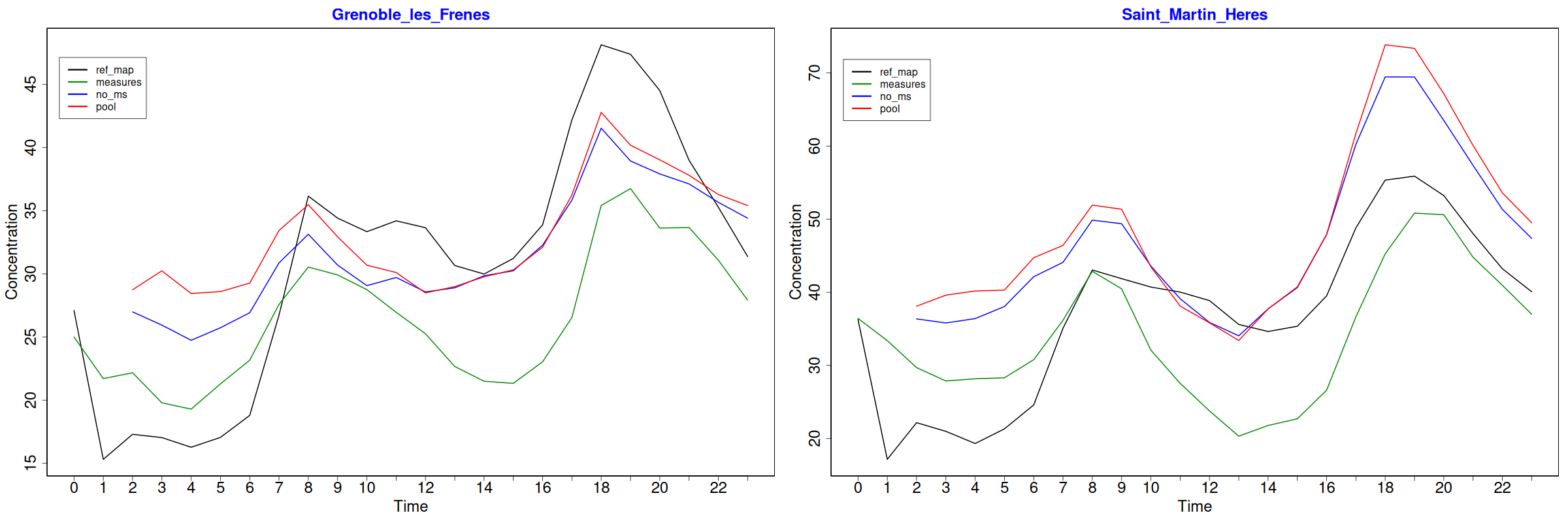} 
   \caption{February average of the estimated concentration for the zone corresponding to the reference stations at Grenoble les Frênes (left) and Saint-Martin-d'Hères (right), as a function of the time of day. In black: the initial map; in green: measurements at the reference station; in violet: ms\_as\_sta; in red: pool; in blue: no\_ms.}
  \label{fig:FreneSta_moyen}
\end{figure}

For each reference station we also plot the estimated concentrations for a given day (Figure \ref{fig:gre_bd_jour}) as well as the estimated average concentration over the February period (Figure \ref{fig:FreneSta_moyen}). This corrected concentrations, shown in red (pool) and blue (no\_ms), are to be compared with the curve on the initial map (in black) and that from the measurements (in green).

As it was first noticed from Figure \ref{fig:moycarte_jour}, the two methods show similar corrections. From Figure \ref{fig:gre_bd_jour}, we see that between 2 am and 8 am, and at the end of the day, the corrected maps tend to act as the initial map. However, between 9 am and 6 pm, both of the corrected maps do better than the initial map, as their curves are close the measurements of the reference station. On this example, our proposed methods thus seem to correct the initial map better when NO$_2$ concentrations are higher.

Figure \ref{fig:FreneSta_moyen} shows two different behaviours that are quite representative of what we see at the other stations. For les Frênes station, the value of the corrected average concentrations are mainly between the concentration obtained from the physicochemical model and the average measured concentrations, whereas for Saint-Martin d'Hères station the corrected concentration is mainly greater than measurements and the initial map. Therefore, it seems that the proposed correction methods do not perform similarly from one zone to another.

\subsection{Correction and micro-sensors}

\begin{figure}[H]
\centering
   \includegraphics[width=0.8\textwidth]{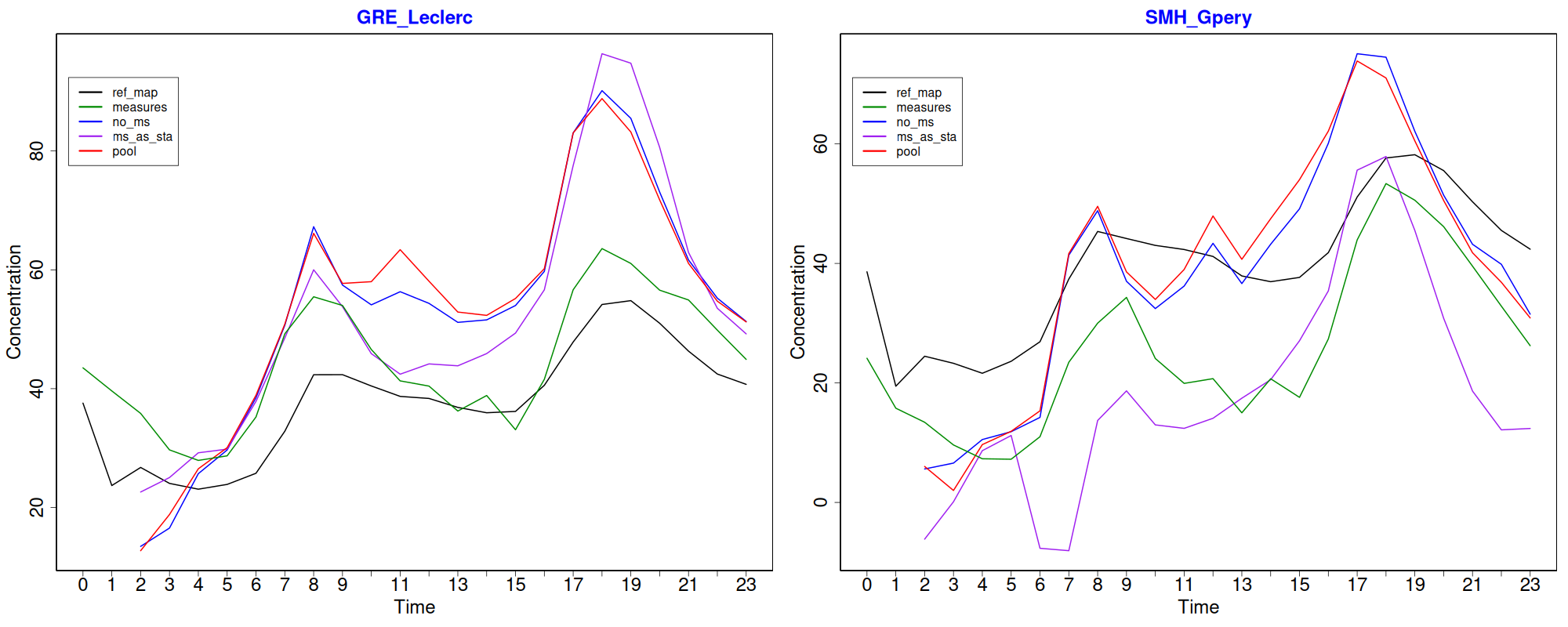}     
  \caption{Left: February average of the estimated concentration at the GRE\_Leclerc micro-sensor as a function of time. In black: the initial map; in green: the micro-sensor measurements;
  respectively in red, blue and purple: the models pool, no\_ms and ms\_as\_sta.
  Right: same for SMH\_Gpery.}
  \label{fig:mc_lec}
\end{figure}

In Figure \ref{fig:mc_lec} we plot the same curves as in Figure \ref{fig:FreneSta_moyen}, but this time for two micro-sensors. In black we give the averaged concentration estimated by the initial map, in green the averaged (uncorrected) measurements made by the micro-sensor, and the other curves correspond to our methods. In blue we represent the estimated concentrations using only the reference devices (no\_ms), in purple the concentration estimated using the corrected sensors as reference stations, and in red the concentration estimated using the pooling method. As it was noticed from Figure \ref{fig:FreneSta_moyen}, the curve of the no\_ms and pool method are very close. However, the method considering sensors as stations leads to different concentrations. At the GRE\_Leclerc sensor, the concentration estimated by the method is usually between the measurements and the estimation by the other method, the only exception being around 6 am. The behaviour of the different methods is quite different at the SMH\_Gpery sensor. Indeed, our estimations do not seem to follow the highs and downs of the initial map. Moreover, we see that the method ms\_as\_sta leads to some inconsistent estimations, since some concentrations (averaged over February) happen to be negative, for example at 6 am and 7 am.

\begin{figure}[H]
  \centering
  \includegraphics[width=0.8\textwidth]{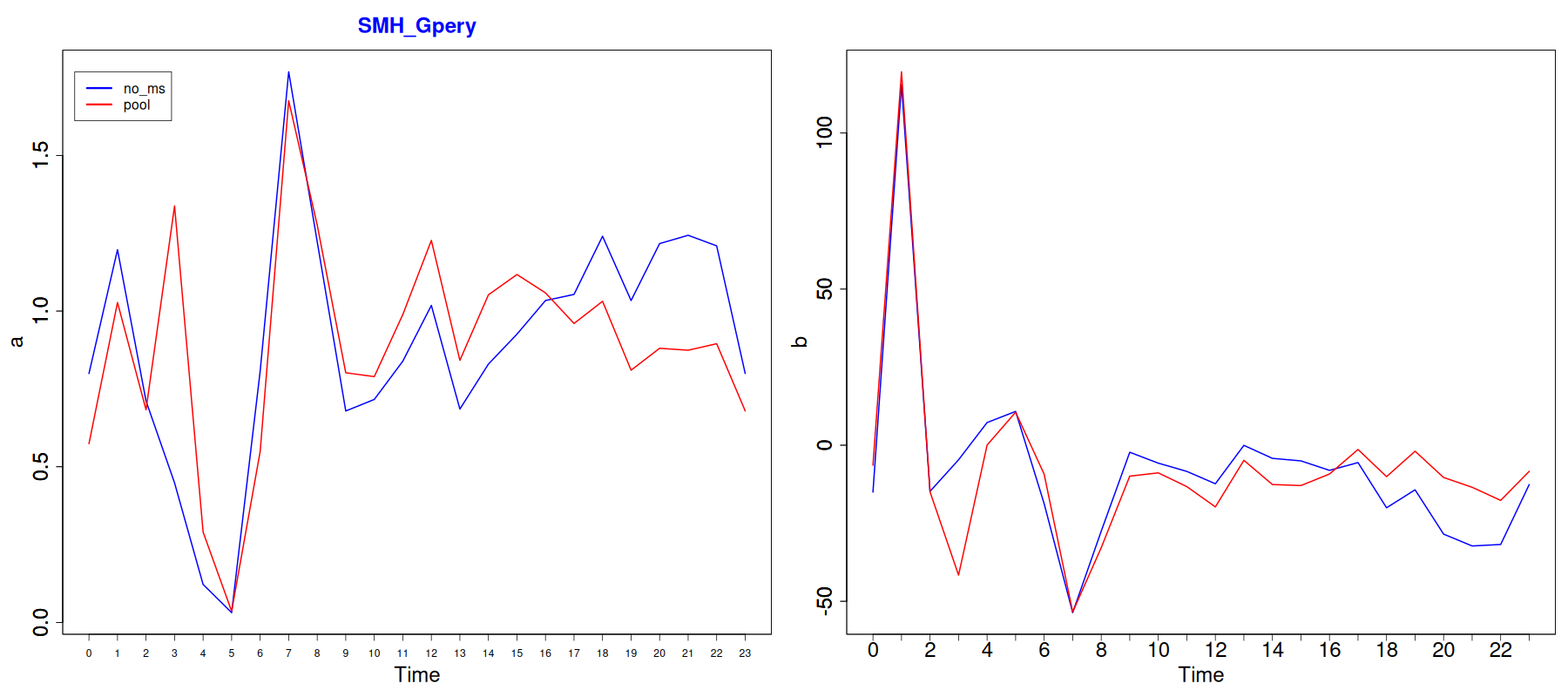}     
  \caption{Correction parameters estimated on the SMH\_Gpery micro-sensor, for each hour. Coefficient $\alpha$ on the left and $\beta$ on the right. In red: pool model; in blue: no\_ms model.}
  \label{fig:ab_SMH_Gpery}
\end{figure}

Figure \ref{fig:ab_SMH_Gpery} shows, for each hour, the value of the parameters optimized by linear regression to correct the SMH\_Gpery micro-sensor measurements. The peak on $\beta$ at 1 am corresponds to an anomaly observed on the reference map, whose values are almost constant at this time (about 11 \mgm3). We note that these parameters show a rather high degree of variability, which does not show a clear temporal pattern and is not reflected in the daily mean concentration dynamics shown in Figure \ref{fig:mc_lec}. This last point is reflected in a relationship between $\alpha$ and $\beta$ that seems to be close to an affine relationship.

\subsection{Validation of the methods}

\subsubsection{A first look at the RMSE}

\begin{figure}[H]
    \centering
    \includegraphics[width = 0.9\textwidth]{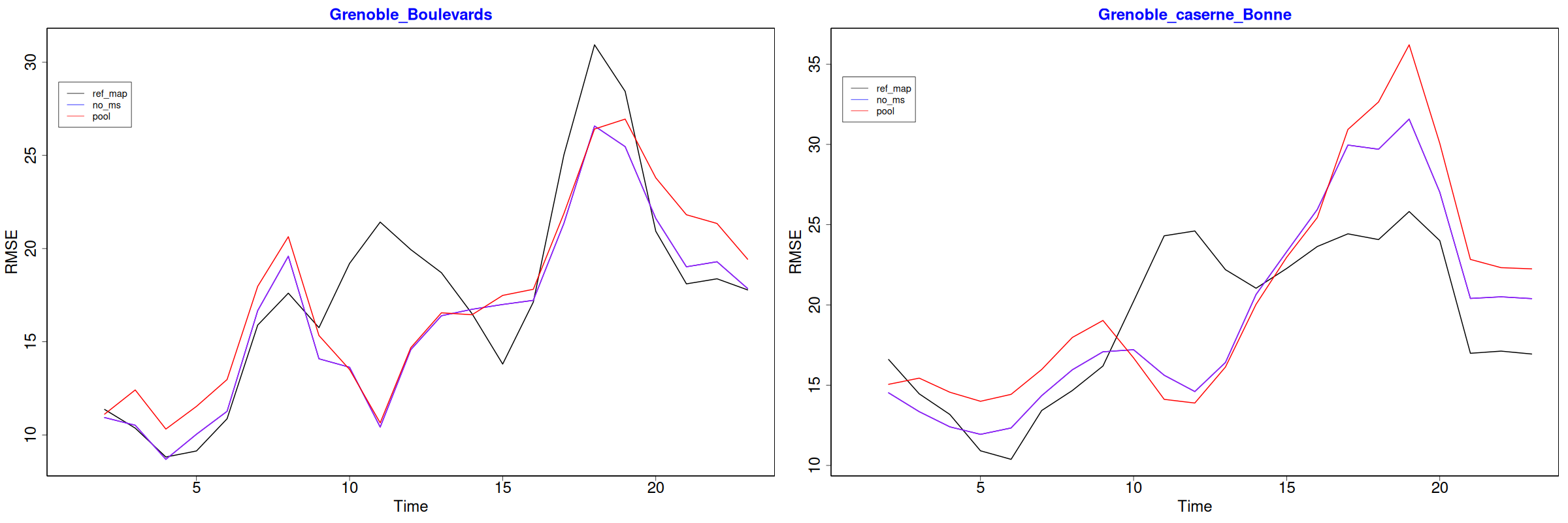}
    \caption{RMSE as a function of the hour of the day for stations Grenoble Boulevards (on the left) and Caserne Bonne (on the right). In black: RMSE of the initial map, in red RMSE of the pooling model, in purple the RMSE of the model ms\_as\_sta (same as no\_ms in this case).}
    \label{fig:rmse_by_hour}
\end{figure}

In order to give a first assessment of the methods and compare them with respect to the initial map, we propose to use the root mean square error (RMSE). Using the notation $X^{i,1}(t)$ for the measurement of reference station $i$ at time $t$ and $\hat{P}(i,t)$ the map of interest (initial, pool, no\_ms or ms\_as\_sta), the RMSE is defined by:

$$RMSE(\hat{P}) = \sqrt{\frac{1}{I\times T}\sum_{i = 1}^{I}\sum_{t = 1}^{T}\left(X^{i,1}(t) - \hat{P}(i,t)\right)^2}$$

where $I$ is the number of reference stations. Note that the computation of RMSE only takes into account the measurements made by reference stations, since they are not biased (in contrary to those of micro-sensors). Moreover, since the coefficients $\rho$ and $C$ are estimated at each hour, it is interesting to compute the RMSE by hour and for the four maps. The temporal RMSEs are given in Figure \ref{fig:rmse_by_hour} for stations Grenoble Boulevards (on the left) and Caserne Bonne (on the right). This figure shows that the corrected map is generally better than the original map, as expected. However, towards the end of the day it is not so clear; sometimes the original map even slightly outperforms our correction. Both models (pool and ms\_as\_sta) show very similar results.

In addition, to assess the overall quality of the procedure, we calculate a cross-validated RMSE using rolling hold-out on reference stations.

\subsubsection{Cross-validation error}
A first numerical evaluation of the interest of the procedure is realized by estimating the error of reconstruction of the concentrations of the reference network, by cross-validation. More precisely, we implement a leave-one-out procedure. For each point, we apply the whole procedure using only the measurements provided by all the other sensors (network 2 microsensors and other network 1 stations) and we can deduce a prediction error at that point.
This process is repeated for all the reference stations and the quality is summarised by the average of the prediction errors. We use all the January observations (at a given) hour to adapt the structure of the model and to fit the model parameters. The performance is evaluated on the days of February, at the same hour.

More precisely, we build 4 models corresponding to hours $h=$ 6am, 9am, 3pm and 6pm and calculate the cross-validation error $$ Err(h)=\sum_{i=1}^6Err^{(-i)}(h)$$

where $ Err^{(-i)}(h)$ is the RMSE over the days of February omitting the $i$th reference  station. 

This gives a global performance (average value of the RMSE over all available hours of February) of about 12 to 20 \mgm3. This evaluation is of quite good quality (12 for urban and rural stations and 20 for traffic ones, after a corrected pairing mode taking into account the nature of the jackknifed station (traffic, urban and rural). Indeed the cross-validation scheme is conservative, as we were faced with a very sparse network 1 and omitting 1 out of 6 could have a significant impact.

\begin{figure}[H]
    \centering
    \includegraphics[width = 0.9\textwidth]{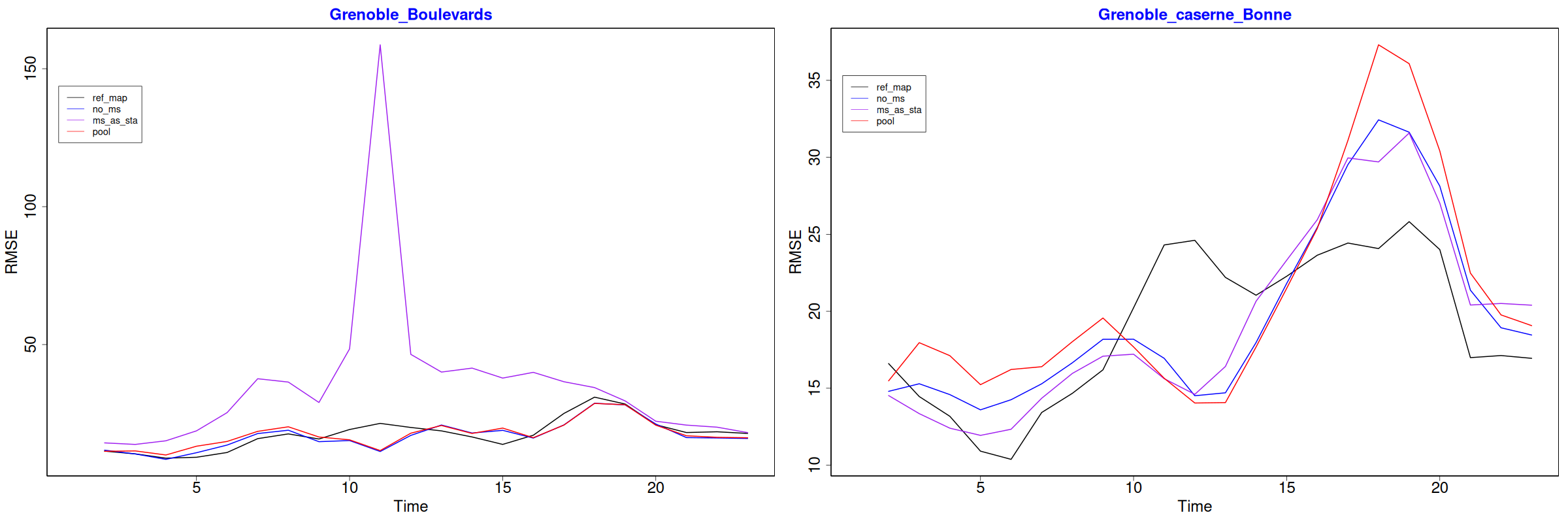}
    \caption{Cross-validated RMSE as a function of the hour of the day for stations Grenoble Boulevards (on the left) and Caserne Bonne (on the right). In black: RMSE of the initial map, in red the RMSE of the pooling model, in purple the RMSE of the model ms\_as\_sta, in blue the RMSE of no\_ms.}
    \label{fig:rmse_by_hour_cv}
\end{figure}

Figure \ref{fig:rmse_by_hour_cv} presents the cross-validation RMSE at two monitoring stations, by hour and by method, computed over February. It appears from both site that our models correct the initial map between 10 am and 3 pm (difference of 10 \mgm3), but fails to reduce the errors at other moments. The figure also illustrates the difference between the models computed with micro-sensors as reference stations and the model no\_ms. Indeed, at station Grenoble Boulevard, we see that the curve of RMSE by hour of the day reaches a peak at 11 am, where the RMSE (150 \mgm3) is significantly different from the other hours. This is due to the fact that for this specific hour, the measurements of the nearest station (which is in that case a microsensor) behave as a white noise leading to an almost null slope in the linear fitting.

\begin{figure}[H]
\centering
   \includegraphics[scale=0.15]{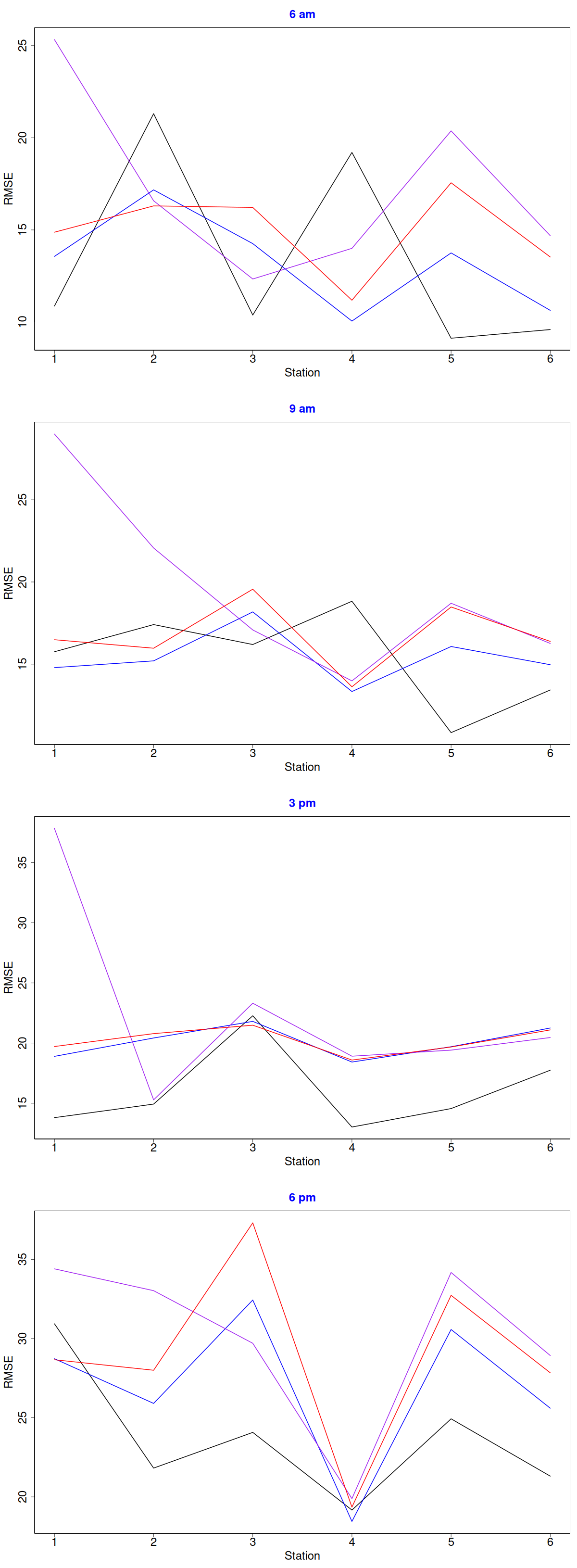}
  \caption{RMSE (computed on February) as a function of the stations, for the different hours (6 am, 9 am, 3 pm and 6 pm). We select the nearest station of the same type for all models: no\_ms in blue, ms\_as\_sta in purple and pool in red. Stations from index 1 to 6 are respectively: Grenoble\_Boulevards, Grenoble\_Rocade\_Sud, Grenoble\_caserne\_Bonne, Fontaine\_les\_Balmes, Grenoble\_les\_Frenes, Saint\_Martin\_Heres.}
\label{fig:rmse_by_station_cv}
\end{figure}

Figure \ref{fig:rmse_by_station_cv} shows the results with the same color code. Since in this case no\_ms and ms\_as\_sta behave differently (because the closest neighbor of a station might be a microsensor), a third curve is shown in blue corresponding to no\_ms. Figure \ref{fig:rmse_by_station_cv} gives the evolution of the performance for the different hours of the day. We plot the RMSE as a function of the stations, for the different hours (6 am, 9 am, 3 pm and 6 pm) for February (the testing period). In each of the 4 plots, there are four curves, corresponding to the initial map and the three proposed models. The NO$_2$ estimated at the jackknife station has been computed using the nearest device of the same type. This device is always a reference station for the model no\_ms (in blue), but can be a micro-sensor with the other methods (ms\_as\_sta in purple, pool in red). For every hour, the curves in blue and red are quite similar, which is consistent with other results. However, the RMSE of the no\_ms method is almost always smaller than the one of the pool method. On the other hand, the method ms\_as\_sta seems to perform very differently from one station to another. For example, the method leads to high RMSE for the station 1 (Grenoble Boulevard), being greater than the other method by 5 to 20 $\mu g / m ^3$; while it is better for the station 3 (Grenoble Caserne). For station 2 (Grenoble Rocade Sud) and 4 (Fontaine-les-Balmes), the cross-validation estimation is sometimes better than the initial map.

\begin{table}[H]
    \centering
    \begin{tabular}{c|c|c|c}
         Station & no\_ms & ms\_as\_sta & pooling \\
         \hline
         Grenoble Boulevards & \textbf{17.3} & 44.7 & 17.8\\
         Grenoble Rocade & \textbf{18.8} & 38.7 & 20.0\\
         Grenoble Caserne & 20.4 & \textbf{20.1} & 21.7\\
         Fontaine les Balmes & \textbf{15.1} & 16.3 & 15.8\\
         Grenoble Frenes & \textbf{20.0} & 21.6 & 21.7\\
         Saint-Martin d'Hères & \textbf{18.7} & 19.3 & 19.7
    \end{tabular}
    
    \caption{(Cross-validated) RMSE computed by each method, using for each station the parameters of the nearest station of the same type. In bold, the best score obtained at the station (not considering reference map).}
    \label{tab:RMSE_by_model}
\end{table}

Table \ref{tab:RMSE_by_model} displays the cross-validation RMSE for each model at each reference station. It appears that the model no\_ms using only reference stations performs slightly better than the others, having the lowest RMSE at 5 stations over 6. The pooling model has similar RMSE, and the model ms\_as\_sta leads to some large RMSE (around 40 \mgm3), due to the specific phenomenon at 11 am previously analyzed (see Figure \ref{fig:rmse_by_hour_cv}, the outlier at 11 am increasing the RMSE).

\subsection{A point and click tool}

\begin{figure}[H]
   \includegraphics[width=\textwidth]{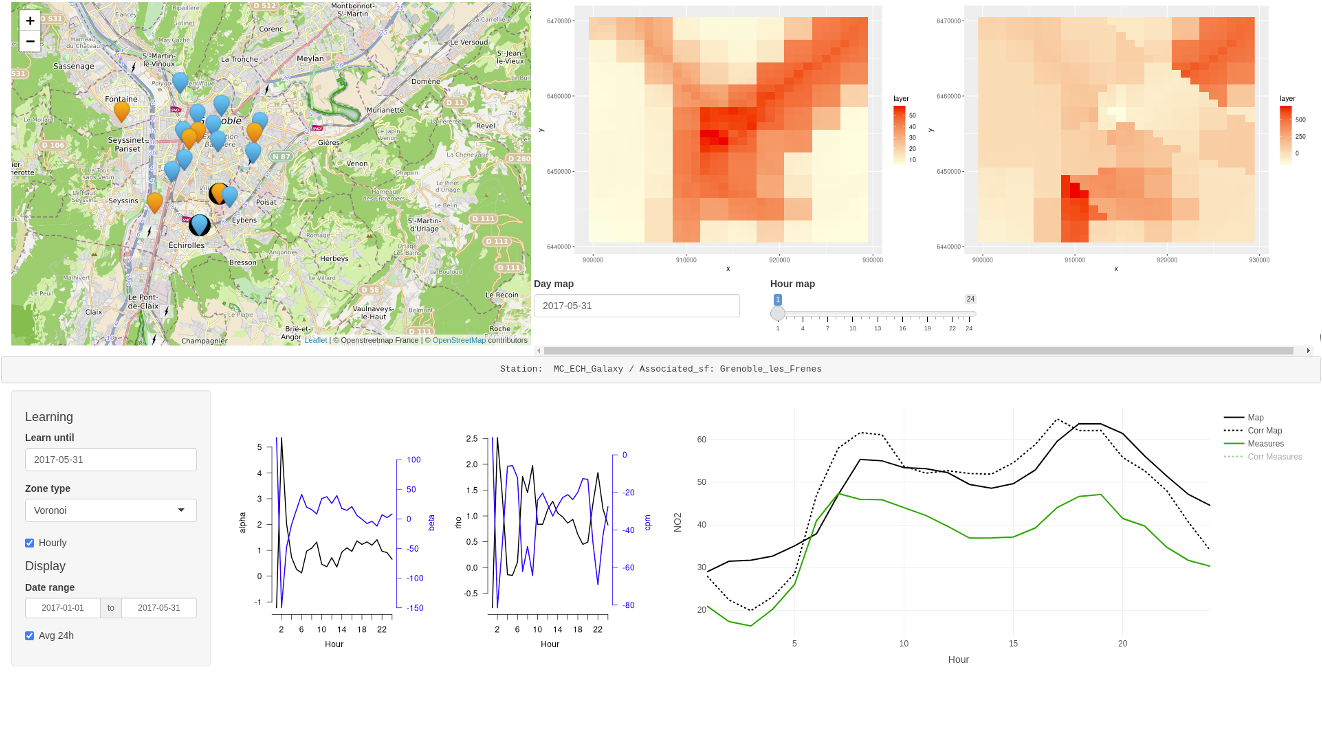}     
  \caption{Point and click tool. A click on reference station/ micro-sensor, on the left, updates the parameters $\rho$ and $C$ $\alpha$ and $\beta$ - displayed at bottom left, the measurement update at bottom right as well as the corrected map. Learn/Estimate until "learn until" (bottom left), hour by hour or globally, displayed in 24h averaging or not.
}
\label{fig:interface}
\end{figure}

In the applied context, we have developed a point and click tool (see Figure \ref{fig:interface}) allowing to visualize the maps (initial and corrected) and measurements, to select the instants considered as the learning period, to estimate and plot the estimated parameters and finally to visualize the corrected map at given time and the corrected measurements over time.  

This tool is currently used in Atmo Normandie, to explore new micro-sensor data (see \cite{thulliez2023air}) and would be publicly available in the future with some additional data.

\section{Perspectives} \label{sec:modele+general}

In this section we collect the prospects that can be deduced from this initial exploration. 

A first group is straightforward. We could explore other spatial partitions, in particular one that takes into account the distance from the city centre or major roads, and explore other temporal breakdowns of the parameters in order to define and implement more validation procedures. This will also allow us to quantify the contribution of micro-sensors to this approach.

Longer term prospects could include taking into account several aspects of modelling that might be of interest to us and that are neglected in the first model presented above:

\begin{itemize}
    \item Firstly, as mentioned above, the noise contained in the micro-sensor measurements may follow a more general distribution or a non-Gaussian distribution.
    \item Similarly, the transformation functions of the concentrations from the micro-sensor measurements can be more general than affine functions. For example, they can depend also on the brand of the microsensor. This choice can be determined according to the first results obtained in the estimation of the $\alpha_j$ and $\beta_j$ parameters in the previous section.
    \item The temporal aspects of the problem have been neglected in the previous model, although they can be integrated on several levels. Firstly, the time range for which we have data can be divided into several periods, which will allow us to estimate the time-dependent parameters $\alpha_{j,t}$ and $\beta_{j,t}$. This generalisation will give an idea of the extent to which the behaviour of the micro-sensors changes over time. Furthermore, according to the results of \cite{soulhac2017model}, it seems relevant to consider the seasonality and the daily rhythm in the modelling of the bias $B_s(t)$. A spatio-temporal modeling of the bias could also be considered, e.g. inspired by \cite{fasso2007air}.
\end{itemize}

\section*{Acknowledgements}

This work takes place in a scientific collaboration between Atmo Normandie, Paris-Saclay University and INSA Rouen Normandie. We would like to thank Véronique Delmas and Michel Bobbia, from Atmo Normandie, for providing the problem and for supporting the statistical study. 

We would like to thank Atmo Auvergne-Rhône-Alpes, the network of the Grenoble area, for kindly authorize and make possible the use of the real data from the Mobicit'air project of Atmo Auvergne-Rhône-Alpes in partnership with Grenoble-Alpes-Métropole, as well for sharing some useful knowledge about these data.

\section*{Data availability statement}
Research data are the property of a third party and are not shared.

\section*{Authors contribution}
All the authors contribute equally to the paper.

\section*{Funding}
This research received no specific grant from any funding agency in the public, commercial, or not-for-profit sectors.

\section*{Disclosure statement}
The authors declare that they have no known competing financial interests or personal relationships that could have appeared to influence the work reported in this paper.

\section*{Ethical Statement}
The authors affirm that this work does not involve experiments on humans or animals.

\bibliographystyle{tfcad}
\bibliography{biblio}

\end{document}